\documentclass[doublecol]{epl2}

\usepackage{amsmath}
\usepackage[margin=0.7in]{geometry}
\newcommand{\bnabla}{\mbox{\boldmath$\nabla$}}

\title{Model of metameric locomotion in active directional filaments}
\author{G. Du\inst{1,2} \and S. Kumari\inst{2} \and F. Ye\inst{1,2,4,5} \and R.
  Podgornik\inst{1,3,4}\thanks{E-mail: \email{rudolfpodgornik@ucas.ac.cn}. Also affiliated with Department of Physics, Faculty of Mathematics and Physics, University of Ljubljana, 1000 Ljubljana, Slovenia.}}
\shortauthor{G. Du\etal}
\institute{\inst{1}Wenzhou Institute, University of Chinese Academy of Sciences, Wenzhou, Zhejiang 325001, China \\ \inst{2}School of Physical Sciences, University of Chinese Academy of Sciences, Beijing 100049, China\\ \inst{3}School of Physical Sciences and Kavli Institute for Theoretical Sciences, University of Chinese Academy of Sciences, Beijing 100049, China\\ \inst{4}Beijing National Laboratory for Condensed Matter Physics and Laboratory of Soft Matter Physics, Institute of Physics, Chinese Academy of Sciences, Beijing 100190, China \\ \inst{5}Oujiang Laboratory, Wenzhou, Zhejiang 325000, China}

\abstract{Locomotion in segmented animals, such as annelids and myriapods (centipedes and millipedes), is generated by a coordinated movement known as {\sl metameric locomotion}, which can be also implemented in robots designed to perform specific tasks. We introduce a theoretical model, based on an active directional motion of the head segment and a passive trailing of the rest of the body segments, in order to formalize and study the metameric locomotion. The model is specifically formulated as a steered Ornstein-Uhlenbeck curvature process, preserving the continuity of the curvature along the whole body filament, and thus supersedes the simple active Brownian model, which would be inapplicable in this case. We obtain the probability density by analytically solving the Fokker-Planck equation pertinent to the model. We also calculate explicitly the correlators, such as the mean-square orientational fluctuations, the orientational correlation function and the mean-square separation between the head and tail segments, both analytically either via the Fokker-Planck equation or directly by either solving analytically or implementing it numerically from the Langevin equations. The analytical and numerical results coincide. Our theoretical model can help understand the locomotion of metameric animals and instruct the design of metameric robots.}
\begin{document}

\maketitle

\section{Introduction}
Active particles are systems far from equilibrium that utilize energy sources to perform non-thermal motions~\cite{vicsek1995,schweitzer1998,ramaswamy2010,menzel2015}. Examples of active particles include both living systems such as cytoskeletal filaments~\cite{julicher2007,gupta2019,foglino2019}, microorganism colonies~\cite{chen2007,tailleur2009}, and bird flocks~\cite{toner2005a}, as well as synthetic systems such as self-propelled colloids~\cite{ballerini2008} and driven granular matter~\cite{deseigne2010a}. Among these active systems one can also include animals with metameric structure that enables them to perform a special type of gait referred to as {\sl metameric locomotion}.
Limbless and many-legged invertebrates can generate a coordinated gait via a collective movement of segments or appendages, not dissimilar to a metachronal wave in cilia arrays~\cite{elgeti2013b}, that efficiently propels the body forward. Inspired by the crawling creatures, robots have been developed that simulate the metameric locomotion in order to perform tasks that otherwise traditional robots would accomplish only inefficiently~\cite{calderon2016,zhan2019}.
While presently the studies of metameric locomotion focus mostly on the experimental observation of living worms and millipedes~\cite{stephens2010,padmanabhan2012,Kuroda2018,Garcia_2020}, as well as related biology inspired robot engineering problems~\cite{aoi2013,Hoffman2013,fang2015,Ishiguro2017,Spinello2017,agostinelli2018,zhan2019}, theoretical models highlighting the dynamical properties of the motion trajectory are lacking. In what follows we formulate an active directional filament model to simulate the metameric locomotion and obtain analytical results describing its properties, including the probability density and physical correlators related to the trajectory of the motion.

We start from the continuity constraint for the filament tracing the body frame as one of the defining features of the model describing metameric locomotion. The shape of the moving filament should be not only continuous in position, but also with continuous orientation and curvature fields, as there are indications that nematodes and other metameric animals trace out locomotion trajectories
with continuous curvature~\cite{padmanabhan2012}, likely related to the fact that the directional steering itself is
continuous, possibly corresponding to lowest energy consumption.
The continuous curvature constraint already disqualifies variants of the active Brownian model~\cite{Pototsky_2012,Cates_2013}, where the orientation of the active particles changes with time continuously but is not differentiable, corresponding to a discontinuous curvature field.
We thus resort to a higher order model dynamics, with continuous position, orientation as well as curvature fields.
Without any sensorial feedback the metameric locomotion can be presumed to persist as rectilinear, however,
the constant influx of environment information and the inaccuracy in the coordinated response can be listed as sources of the unavoidable curvature noise. We therefore in addition assume that the curvature can be reasonably described as undergoing an  Ornstein-Uhlenbeck process, possibly steered by external forces. In addition, due to the connectivity of the metameric body the locomotion shows a trailing property, where distal body sections follow the trajectory of the head.
For the same reason the body filament can be considered as inextensible globally as well as locally,
leading to a filament velocity with a constant norm. All the listed constraints of the metameric locomotion model can be subsumed by describing it as an {\sl active directional filament motion}, generated by an active head with curvature noise and a passive body trailing the trajectory of the active head.

The model constraints laid out lead to a set of equations of a higher order Langevin form that allows us to obtain the probability density of and the correlators characterizing the metameric locomotion analytically.
Two paths are then taken to solve the model: (i) obtain the correlators, {\sl i.e.}, the mean-square orientational fluctuations, the orientational correlation function and the mean-square separation between the head and tail of the body filament, directly from the solution of the Langevin equations; (ii) obtain the correlators via the solution of the corresponding Fokker-Planck equation, which gives the probability density of the head particle, from which one can calculate the probability density for the whole moving  filament. Since the solution of the Fokker-Planck equation carries the complete information regarding the metameric locomotion, we will mainly pursue  the calculation of the second path, but also establish that the results from the Fokker-Planck method and those stemming directly from the Langevin equations coincide. Moreover, we will numerically integrate the Langevin equations and obtain the running averages of the trajectories and show that they agree well with the analytical results, consequently validating both the analytical as well as the numerical results.

\section{Model}
\begin{figure}[tpb]
\centering
\includegraphics[width=.9\linewidth]{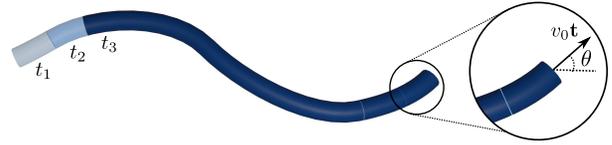}
\caption{\label{fig:schematic}Schematic illustration of the active directional filament model on a two-dimensional infinite flat plane at three different times: $t_1 \leq t_2 \leq t_3$. The active directional filament consists of an active head segment and passive body segments trailing the trajectory of the active head segment. The magnified region defines the velocity, with constant norm $v_0$, tangential vector ${\bf t}$ and the orientational angle $\theta$ of the active head segment.}
\end{figure}

We now propose the analytical form of the active directional filament model for metameric locomotion. The active filament lives on a two-dimensional infinite flat plane. It consists of an active head and a passive trailing body. The motion of the head maintains a velocity with constant norm \(v_{0}\). The length of the filament is \(L\). Denote the position of the body segment at an arclength distance \(l\) from the head as \(\mathbf{r}_{\tilde{l}}(t)\), where \(t\) is time and the dimensionless arclength is defined as  \(\tilde{l} = l/L\) with \( 0 < \tilde l \leq 1\). Then the position of head and tail are \(\mathbf{r}_{0}(t)\) and \(\mathbf{r}_{1}(t)\), respectively.
Denote the time for the filament to move a distance equal to its body length as \(T= L/v_{0}\).
Since the rest of the body segments are trailing the trajectory of the head, this implies that for \(t>T\)
\begin{align}
\mathbf{r}_{\tilde l}(t) = \mathbf{r}_{0}(t-\tilde{l} T).
\end{align}
In the main text, we will focus on the large time regime \(t > T\).
The marginal case with \(t \leq T\) will be trivial to discuss after we obtain the results of the case with \(t > T\)~\cite{suppl}.
As long as no ambiguity rises, we will omit the subscript of the variables pertinent to the active head for simplicity.
The motion of the active head with a constant velocity norm follows the dynamical equations~\cite{castro-villarreal2019}
\begin{align} \label{eq:frenet}
\frac{\mathrm{d}\mathbf{r}(t)}{\mathrm{d}t} &= v_{0} \, \mathbf{t}(t), \\
\frac{\mathrm{d}\mathbf{t}(t)}{\mathrm{d}t} &= v_{0} \kappa(t)\mathbf{n}(t),
\end{align}
where \(\mathbf{t}(t)\) and \(\mathbf{n}(t)\) are, respectively, the tangential and normal vectors of the head motion, while \(\kappa(t)\) is the instantaneous curvature.

In two-dimensional space, the tangential vector can be rewritten as \(\mathbf{t}= (\cos\theta(t),\, \sin\theta(t))\) with \(\theta(t)\) being the orientational angle of the tangential vector of the head segment, see Fig.~\ref{fig:schematic}. Then it follows that
\begin{align} \label{eq:ori}
\frac{\mathrm{d}\theta(t)}{\mathrm{d}t} = v_{0} \kappa(t).
\end{align}
By assumption the curvature of the head segment undergoes a steered Ornstein-Uhlenbeck process
\begin{align} \label{eq:langevin}
  \frac{\mathrm{d}\kappa(t)}{\mathrm{d}t} = -\beta \kappa(t) - \gamma \bnabla U \cdot {\bf n}(t) +  \xi(t),
\end{align}
where \(\beta >0\) is curvature decay constant and \(\xi(t)\) is white noise
\begin{align}
\langle \xi(t)\rangle = 0, \quad \langle\xi(t)\xi(t')\rangle = g \,\delta(t-t'),
\end{align}
with \(g\) being the constant noise amplitude. For generality, we also added an external force term, \(  -\bnabla U \), derived from a position and orientation dependent steering potential \(U({\bf r}, {\bf t})\), with $\gamma$ being a positive coupling coefficient.

From the above Langevin equations, we can derive the corresponding Fokker-Planck equation.
Note that in the absence of the steering potential the correlators related to position \(\mathbf{r}(t)\) can be calculated from the orientation data,
see Sec.~Correlators, so it suffices to solve the Fokker-Planck equation for the
probability density as a function of only orientation and curvature.
 Assume the probability of finding at time \(t\) the active head in the orientation interval \([\theta,\, \theta+\mathrm{d}\theta]\) and
 curvature interval \([\kappa, \,\kappa + \mathrm{d}\kappa]\) is \(P_{0}(\theta,\kappa,t | \theta_{0}, \kappa_{0}, 0) \,\mathrm{d}\theta\,\mathrm{d}\kappa\), where \(\theta\) and \(\kappa\) are, respectively, the orientation and curvature at time \(t\), and \(\theta_{0}\) and \(\kappa_{0}\) are, respectively, the orientation and curvature at time \(t=0\). Then we end up with the following Fokker-Planck equation
 \begin{align}
\label{eq:kfp}
 \frac{\partial }{\partial t} P_{0}
 = - \frac{\partial}{\partial\theta}\left(v_{0}\kappa P_{0}\right)
+ \frac{\partial}{\partial \kappa} \left(\beta \kappa P_0\right)
+ \frac{g}{2} \frac{\partial^{2}}{\partial \kappa^{2}} P_0,
 \end{align}
 which coincides with a two-dimensional Ornstein-Uhlenbeck process and is thus analytically solvable~\cite{suppl}. This is of course only true when there is no steering potential or the steering potential depends only on orientation~\cite{suppl}.

\section{Probability density}
Assume the active head initially has orientation and curvature distribution \(P_{0}(\theta, \kappa,t| \theta_{0}, \kappa_{0},0)|_{t=0} = \delta(\theta-\theta_{0}) \delta(\kappa-\kappa_{0})\).
Denote \(\mathbf{z}=\begin{pmatrix}\theta & \kappa \end{pmatrix}^{\mathsf{T}}\),
\(\mathbf{z}(t)=\begin{pmatrix}\theta(t) & \kappa(t) \end{pmatrix}^{\mathsf{T}}\),
where \(\theta(t)\) and \(\kappa(t)\) are, respectively, the mean orientation and curvature dependent on time
\begin{align} \label{eq:mean-ori-curv}
\theta(t) = \theta_0 + \frac{1-\mathrm{e}^{-\beta t}}{\beta} v_0\kappa_0, \quad \kappa(t) = \mathrm{e}^{-\beta t} \kappa_0
\end{align}
and \(\sigma(t) = \bigl(\begin{smallmatrix}\sigma_{\theta\theta} & \sigma_{\theta\kappa}\\ \sigma_{\kappa\theta} & \sigma_{\kappa\kappa}\end{smallmatrix}\bigr)\), which is the variance matrix with elements
\begin{subequations}
\begin{align}
\sigma_{\theta\theta} &= \frac{gv_{0}^{2}}{2\beta^{3}} \left( 2\beta t - \mathrm{e}^{-2\beta t} + 4\mathrm{e}^{-\beta t} - 3\right), \\
\sigma_{\theta\kappa} &= \sigma_{\kappa\theta} = \frac{gv_{0}}{2\beta^{2}} \left( \mathrm{e}^{-\beta t} - 1\right)^{2}, \\
\sigma_{\kappa\kappa} &= \frac{g}{2\beta} \left(1 - \mathrm{e}^{-2\beta t}\right).
\end{align}
\end{subequations}
Then the solution of Eq.~\eqref{eq:kfp} is given by~\cite{risken1989,suppl}
\begin{align} \label{eq:prob-2d-ou}
& P_{0}(\theta, \kappa, t | \theta_{0}, \kappa_{0}, 0) = \frac{1}{2\pi \sqrt{\text{det}~ \sigma(t)}} \nonumber\\
& \qquad \times\exp \left\{-\frac{1}{2} \left[\mathbf{z}-\mathbf{z}(t)\right]^{\mathsf{T}}\left[\sigma(t)\right]^{-1}\left[\mathbf{z}-\mathbf{z}(t)\right]\right\}.
\end{align}
Clearly, the probability density depends explicitly on the orientation, curvature and time.
It is noteworthy that the range of orientation in the probability density function is $(-\infty, \infty)$ instead of
being wrapped into the interval $[0, 2\pi)$. The reason is that the state of a filament is not a periodic function of the orientations of individual segments.
Specifically, different from a rod, the filament can coil into spiral structures when
the orientation of the head varies to exceed the range $[0, 2\pi)$.

We now remark on the behavior of the probability density Eq.~\eqref{eq:prob-2d-ou} at large time limit.
When \(t \to \infty\), \(\theta(t)\to \theta_{0} + v_{0}\kappa_{0}/\beta\), \(\kappa(t)\to 0\), \(\sigma_{\theta\theta}\to gv_{0}^{2}(2\beta t -3)/(2\beta^{3})\), \(\sigma_{\theta\kappa} \to gv_{0}/(2\beta^{2})\) and \(\sigma_{\kappa\kappa} \to g/(2\beta)\).
The mean curvature vanishes while the mean orientation remains finite.
The orientational component of the variance matrix diverges linearly with increasing time while
other components remains finite. It indicates a localized distribution of curvature
around zero and an extended distribution of orientation around a finite value at large time.

The probability density of the whole filament, trailing the head motion, can be readily obtained from the probability density of the active head. The probability density of finding a segment of a single filament at time \(t\) with
orientation \(\theta\) and curvature \(\kappa\) is
\begin{align}
P(\theta, \kappa, t) = \int_{0}^{1}\mathrm{d}\tilde{l} ~\Big< \delta(\theta-\theta_{\tilde{l}}(t))\delta(\kappa-\kappa_{\tilde{l}}(t))\Big>_{\xi}.
\end{align}
Recall that the passive segments of the filamentous body follow the trajectory of the active head. The state of the  passive segments therefore lags behind the state of the head, and we conclude that
\begin{align}
\theta_{\tilde{l}}(t) = \theta_{0}(t-\tilde{l}T), \quad
\kappa_{\tilde{l}}(t) = \kappa_{0}(t-\tilde{l}T).
\end{align}
Note that \(P_{0}(\theta, \kappa, t) = \langle\delta(\theta-\theta_{0}(t))\delta(\kappa-\kappa_{0}(t))\rangle_{\xi}\).
Then after substitution of the integral variable
\begin{align}
P(\theta, \kappa, t) = \frac{1}{T} \int_{t-T}^{t}\mathrm{d}t'P_{0}(\theta, \kappa, t') \quad (t>T).
\end{align}
The above integral can be performed numerically.
Let us define the dimensionless variables by using the following characteristic quantities:
we define the characteristic time \(t_{c}= 1/\beta\), length \(l_{c}= v_{0}/\beta\), curvature \(\kappa_{c}= (g/\beta)^{1/2}\) and orientational angle \(\theta_{c} = (gv_{0}^{2}/ \beta^{3})^{1/2}\), yielding the dimensionless time \(\bar{t}= t/ t_{c}\), length \(\bar{l} = l/l_{c}\), curvature \(\bar{\kappa} = \kappa/\kappa_{c}\), orientational angle \(\bar{\theta} = \theta/\theta_{c}\) and noise \(\bar{\xi} = \xi t_{c}/ \kappa_{c}\).

\begin{figure}[tbp]
\centering
\includegraphics[width=.9\linewidth]{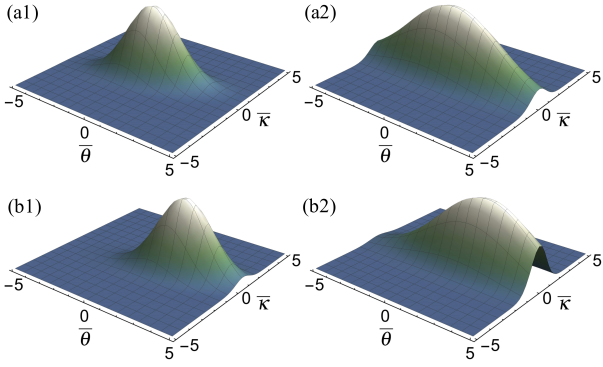}
\caption{\label{fig:prob_density}Plots of the probability densities of finding a segment of a single filament with orientation \(\bar{\theta}\) and curvature \(\bar{\kappa}\) at different times with varying initial orientations and curvatures.
The time \(\bar{t} = 4\) in panels (a1, b1), while \(\bar{t} = 10\) in panels (a2, b2).
The initial orientation \(\bar{\theta}_{0} = 0\) and curvature \(\bar{\kappa}_{0} = 0\) in panels (a1, a2), while \(\bar{\theta}_{0} = 1\) and \(\bar{\kappa}_{0} = 1\) in panels (b1, b2).
In all the subfigures the length of the filament \(\bar{L} = 1\).
}
\end{figure}

Shown in Fig.~\ref{fig:prob_density} are the probability densities of finding a segment of a single filament dependent on orientation and curvature of the segment at different times with varying initial orientations and curvatures.
The time \(\bar{t} = 4\) in Fig.~\ref{fig:prob_density}(a1, b1), while \(\bar{t} = 10\) in Fig.~\ref{fig:prob_density}(a2, b2).
The initial orientation \(\bar{\theta}_{0} = 0\) and curvature \(\bar{\kappa}_{0} = 0\) in Fig.~\ref{fig:prob_density}(a1, a2), while \(\bar{\theta}_{0} = 1\) and \(\bar{\kappa}_{0} = 1\) in Fig.~\ref{fig:prob_density}(b1, b2).
The length of the chain is fixed to be \(\bar{L} = 1\) in all the subfigures.
The probability density is partially heterogeneous and localized at time \(\bar{t}=4\) [Fig.~\ref{fig:prob_density} (a1)] and becomes completely heterogeneous at a larger time \(\bar{t}=10\) [Fig.~\ref{fig:prob_density} (a2)]. At large times, due to the diverging orientational component and finite curvature component in the variance matrix, the distribution of orientation is extended, while the distribution of curvature remains localized. Different initial orientation and curvature [Fig.~\ref{fig:prob_density}(b1, b2)] affect only the positions of the mean orientation and curvature. The shapes of the distributions are, however, not altered.

\section{Correlators}

Now that we have analytically solved the Fokker-Planck equation and obtained the probability density, we can calculate correlators such as mean-square orientational fluctuations, orientational correlation function and mean-square separation. The correlators depend in general on two parameters: time as well as position along the filament.

We first discuss generally the correlators of the active head at time \(t_{1}\) and \(t_{2}\).
Since the state of the distal segments of the filament lags behind the state of the active head, the correlators between the head and tail at time \(t_2\) are just a special case with \(t_{1}= t_{2} -T\) if \(t_2 > t_1\).
The mean-square separation is related to the orientational correlation function by
 \begin{align}
\langle\left(\mathbf{r}(t)-\mathbf{r}(t_{0})\right)^{2}\rangle_{\xi}
= v_{0}^{2} \int_{t_{0}}^{t} \mathrm{d}t_{1} \int_{t_{0}}^{t} \mathrm{d} t_{2} \langle\mathbf{t}(t_{1})\cdot\mathbf{t}(t_{2})\rangle_{\xi},
\end{align}
while the orientational correlation function can be calculated from the mean-square orientational fluctuations.
Note first that
\(\langle\mathbf{t}(t_{1})\cdot \mathbf{t}(t_{2}) \rangle_{\xi} = \langle\cos(\theta_{2} - \theta_{1})\rangle_{\xi}\)
and
\(\theta_2-\theta_1 = v_{0} \int_{t_{1}}^{t_{2}} \kappa(t')\mathrm{d} t'\),
where \(\theta_{2}\equiv\theta(t_{2})\) and \(\theta_{1}\equiv\theta(t_{1})\).
Since \(\kappa\) undergoes the Ornstein-Uhlenbeck process, \(\theta_{2}-\theta_{1}\) is a Gaussian variable whose characteristic function can be readily obtained, and the orientational correlation function \(\langle\mathbf{t}(t_{1})\cdot \mathbf{t}(t_{2}) \rangle_{\xi}\) follows as just the real part of the characteristic function of \(\theta(t_{2})-\theta(t_{1})\), that is
\begin{align}
\langle\mathbf{t}(t_{1})\cdot \mathbf{t}(t_{2}) \rangle_{\xi} =& \,
\exp\left\{-\frac{1}{2} \left[\langle(\theta_{2} - \theta_{1})^{2}\rangle_{\xi}-\langle\theta_{2} - \theta_{1}\rangle_{\xi}^2\right] \right\} \nonumber\\
&\times\cos\langle\theta_{2} - \theta_{1}\rangle_{\xi}.
\end{align}
The calculations of the mean-square separation and the orientational correlation function then reduce to the calculations of the mean and the mean-square orientational fluctuations, which can be obtained either from the Fokker-Planck probability density or directly from the Langevin equations.

\begin{figure*}[tbp]
\centering
\includegraphics[width=.9\linewidth]{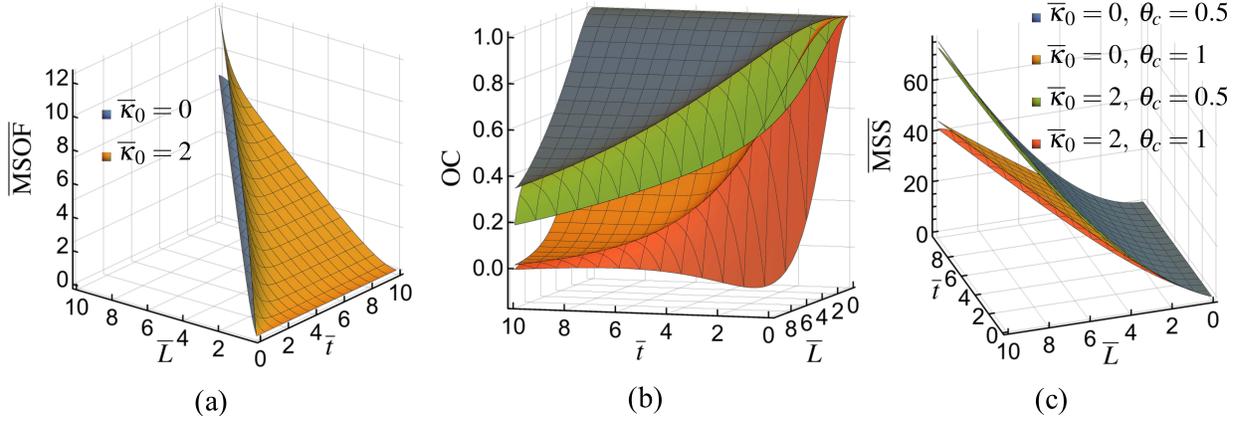}
\caption{\label{fig:correlators_3D}3D plots of correlators between the head segment and the tail segment of a filament obtained from analytical calculations as functions of both time and the length of the filament.
  (a) Mean-square orientational fluctuations (MSOF); (b) orientational correlation (OC) function; (c) mean-square separation (MSS).
  Both the mean-square orientational fluctuations and the mean-square separation are reduced in dimensionless forms.
  The legend in (b) is the same as that in (c).
}
\end{figure*}

We first calculate the mean and the mean-square orientational fluctuations from the probability density.
The mean value can be expressed as
\(
  \langle\theta_{2} - \theta_{1}\rangle_{\xi} = \int\mathrm{d} \theta_{2}\mathrm{d}\kappa_{2} \mathrm{d}\theta_{1} \mathrm{d}\kappa_{1} (\theta_2\!-\theta_{1}) P_{0}(\theta_{2}, \kappa_{2}, t_{2}; \theta_{1}, \kappa_{1}, t_{1}),
\)
where \(P_{0}(\theta_{2}, \kappa_{2}, t_{2}; \theta_{1}, \kappa_{1}, t_{1})\) is the joint probability density (assuming \(t_{1} < t_{2}\))
\(
  P_{0}(\theta_{2}, \kappa_{2}, t_{2}; \theta_{1}, \kappa_{1}, t_{1})
  = P_{0}(\theta_{2}, \kappa_{2}, t_{2}| \theta_{1},\kappa_{1}, t_{1}) P_{0}(\theta_{1}, \kappa_{1}, t_{1})
\).
Note that  \(P_{0}(\theta_{1}, \kappa_{1}, t_{1})\!=\) \(\!P_{0}(\theta_{1}, \kappa_{1}, t_{1}\,|\, \theta_{0}, \kappa_{0}, 0)\) so that
\begin{align} \label{eq:meanr}
\langle\theta_{2} - \theta_{1}\rangle_{\xi} = \frac{v_{0}\kappa_{0}}{\beta} \left(\mathrm{e}^{-\beta t_{1}} - \mathrm{e}^{-\beta t_{2}}\right).
\end{align}
Likewise, we obtain the mean-square orientational fluctuations
\begin{align} \label{eq:msr}
\langle(\theta_{2} - \theta_{1})^{2}\rangle_{\xi}
=&\, \frac{gv_{0}^{2}}{\beta^{3}} \left(\beta |t_{2}-t_{1}| - 1 + \mathrm{e}^{-\beta|t_{2}-t_{1}|}\right) \nonumber\\
&+ \frac{v_{0}^{2}(\kappa_{0}^{2}\beta -g/2)}{\beta^{3}} \left(\mathrm{e}^{-\beta t_{2}} - \mathrm{e}^{-\beta t_{1}}\right)^{2}.
\end{align}
From here it follows that for \(t>T\) the head-to-tail mean-square orientational fluctuations at time \(t\) is
\begin{align} \label{eq:msr-ht}
\langle(\theta(t) - &\theta(t-T))^{2}\rangle_{\xi}
=\, \frac{gv_{0}^{2}}{\beta^{3}} \left(\beta \,T - 1 + \mathrm{e}^{-\beta \,T}\right) \nonumber\\
&+ \frac{v_{0}^{2}(\kappa_{0}^{2}\beta -g/2)}{\beta^{3}} \left(\mathrm{e}^{-\beta t} - \mathrm{e}^{-\beta (t-T)}\right)^{2}.
\end{align}
At large time \(t\), if the length of the chain \(L\) is large, an effective persistence length can be defined as \( l_{\mathrm{p}} = 2{\beta^{2}}/{gv_{0}}, \) which depends quadratically on curvature decay constant \(\beta\) and inversely on the noise amplitude \(g\) and the norm of velocity \(v_{0}\), so that in this limit the active directional filament behaves as an effective worm-like filament.

From the Langevin equation Eq.~\eqref{eq:langevin}, we can obtain the same expressions for the mean and the mean-square orientational fluctuations as
Eqs.~(\ref{eq:meanr}, \ref{eq:msr}).
From Eq.~\eqref{eq:langevin}, the curvature can be obtained as
\begin{align}
\kappa(t) = \mathrm{e}^{-\beta t} \int_{0}^{t} \xi(t') \mathrm{e}^{\beta t'} \mathrm{d} t' + \kappa_{0} \mathrm{e}^{-\beta t}.
\end{align}
Then the mean value of orientational fluctuations is
\begin{align}
  \langle\theta_{2} - \theta_{1}\rangle_{\xi} = v_0 \int_{t_{1}}^{t_{2}} \langle\kappa(t')\rangle_\xi\mathrm{d}t',
\end{align}
which coincides exactly with Eq.~\eqref{eq:meanr}.
The curvature correlation function is
\begin{align}
\langle\kappa(t_{1})\kappa(t_{2})\rangle_{\xi}
=&\, \frac{g}{2\beta} \left( \mathrm{e}^{-\beta|t_{1} - t_{2}|} - \mathrm{e}^{-\beta(t_{1} + t_{2})}\right) \nonumber\\
&+ \kappa_{0}^{2} \mathrm{e}^{-\beta(t_{1}+t_{2})},
\end{align}
wherefrom
\begin{align}
\langle(\theta_{2}-\theta_{1})^{2}\rangle_{\xi} &= v_{0}^{2} \int_{t_{1}}^{t_{2}} \mathrm{d} t' \int_{t_{1}}^{t_{2}} \mathrm{d} t'' \langle\kappa(t')\kappa(t'')\rangle_{\xi},
\end{align}
which reduces exactly back to Eq.~\eqref{eq:msr}.
We have therefore obtained the same results for the mean and the mean-square orientational fluctuations both from the Fokker-Planck equation as well as from the  Langevin equations.

\section{Numerical results}

To validate the previous analytical results, we numerically integrate Eqs.~(\ref{eq:frenet}, \ref{eq:ori}, \ref{eq:langevin}).
In dimensionless form, Eqs.~(\ref{eq:frenet}, \ref{eq:ori}, \ref{eq:langevin}) become
\begin{align}
  \frac{\mathrm{d}\bar{\mathbf{r}}}{\mathrm{d}\bar{t}} &= \mathbf{t} = (\cos\theta, \sin\theta), \\
\frac{\mathrm{d}\bar{\theta}}{\mathrm{d}\bar{t}} &= \bar{\kappa}, \\
\frac{\mathrm{d}\bar{\kappa}}{\mathrm{d}\bar{t}} &= -\bar{\kappa} + \bar{\xi}
\end{align}
where
\begin{align}
\langle\bar{\xi}(\bar{t})\rangle = 0, \quad \langle \bar{\xi}(\bar{t}_{1}) \bar{\xi}(\bar{t}_{2})\rangle = \delta(\bar{t}_{1} - \bar{t}_{2}).
\end{align}
For the curvature, the following result containing scaled time-transformed Wiener process is used
\begin{align}
\bar{\kappa}(\bar{t}) = \bar{\kappa}_{0} \mathrm{e}^{- \bar{t}} + \frac{1}{\sqrt{2}} \mathrm{e}^{- \bar{t}} W_{\mathrm{e}^{2 \bar{t}} - 1}.
\end{align}
To integrate the orientational angle we use the discretized Euler version of dynamics
\begin{align}
\bar{\theta}(\bar{t}+\mathrm{d}\bar{t}) = \bar{\theta}(\bar{t}) + \frac{1}{2}\left[\bar{\kappa}(\bar{t}+\mathrm{d}\bar{t}) + \bar{\kappa}(\bar{t})\right] \mathrm{d}\bar{t},
\end{align}
just as for the case of the position vector dynamics
\begin{align}
\bar{\mathbf{r}}(\bar{t} + \mathrm{d}\bar{t}) = \bar{\mathbf{r}}(\bar{t}) + \mathbf{t}(\bar{t}) \mathrm{d} \bar{t}.
\end{align}

\begin{figure}[tbp]
\centering
\includegraphics[width=.9\linewidth]{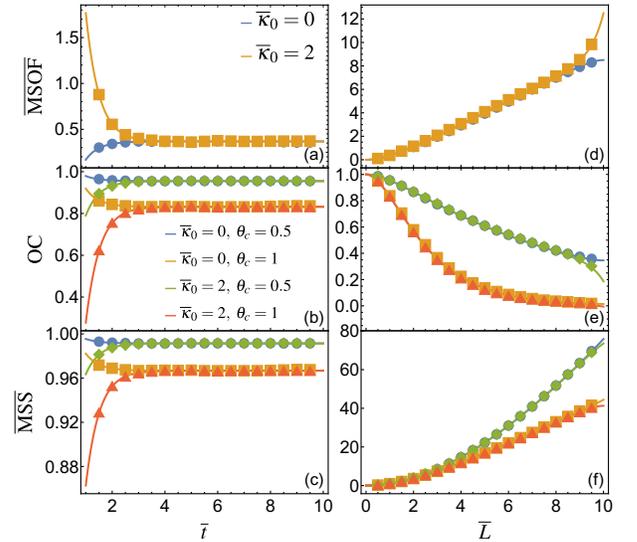}
\caption{\label{fig:correlators_2D}Cross sections of the 3D plots in Fig.~\ref{fig:correlators_3D} at fixed length \(\overline{L}=1\) (a, b, c) and fixed time \(\overline{t}=10\) (d, e, f) obtained both from analytical (solid lines) and numerical (discrete points) calculations.
  The correlators, \emph{i.e.}, the mean-square orientational fluctuations (a, d), the orientational correlation function (b, e) and the mean-square separation (c, f) are all presented in dimensionless form.
  The legend in (d) is identical to that in (a), as the legends in (c, e, f) are identical to that in (b).
  The numerical results obtained by the running averages over \(10^5\) trajectories coincide with the analytical results.
}
\end{figure}

Shown in Fig.~\ref{fig:correlators_3D} are the 3D plots of the correlators, \emph{i.e.}, the mean-square orientational fluctuations [Fig.~\ref{fig:correlators_3D}(a)],
the orientational correlation function [Fig.~\ref{fig:correlators_3D}(b)], and the mean-square separation [Fig.~\ref{fig:correlators_3D}(c)]
between the head segment and the trailing tail segment of the filamentous body.
Fig.~\ref{fig:correlators_2D} displays the cross sections of the 3D plot in Fig.~\ref{fig:correlators_3D} at fixed length \(\bar{L} = 1\) [Fig.~\ref{fig:correlators_2D}(a, b, c)] and at fixed time \(\overline{t} = 10\) [Fig.~\ref{fig:correlators_2D}(d, e, f)].
The solid lines are analytical results obtained by either solving the Fokker-Planck equation or directly calculating from the Langevin equations.
The discrete data are the ones obtained by numerically integrating the Langevin equations.
The numerical results coincide with the analytical ones, consequently validating both the analytical as well as the numerical results.
Note that the mean-square orientational fluctuations Eq.~\eqref{eq:msr-ht} does not depend on initial orientation. The initial curvature only affects
the correlators at small times. At large times, as shown in Fig.~\ref{fig:correlators_2D}(a, b, c), the initial curvature is irrelevant because the exponential factor
in the term containing the initial curvature in Eq.~\eqref{eq:msr-ht} decays to zero at large times.
\(\theta_{c} = (gv_{0}^{2}/ \beta^{3})^{1/2}\) characterizes the orientational fluctuations during the characteristic time \(t_{c} = 1/\beta\). Large curvature diffusion constant \(g/2\) and norm of active velocity \(v_{0}\) both promote the orientational fluctuations, while higher ability to maintain a straight body (large \(\beta\)) suppresses the orientational fluctuations. Consequently, higher \(\theta_{c}\) results in higher mean-square orientational fluctuations, lower orientational correlation, and smaller mean-square separation between the head segment and the tail segment of the filamentous body. From Fig.~\ref{fig:correlators_2D}(f), one can see that the mean-square separation depends quadratically on the length at small length and linearly at large length. At medium lengths, there is a crossover from the ballistic to the diffusive behavior of the mean-square separation.

\section{Conclusions}

In summary, we defined an active directional filament model to describe the metameric locomotion of both segmented animals as well as segmented robots.
As the curvature of these segmented bodies should be continuous, we designed a model with this property explicitly enforced, so that by necessity it generalizes
the active Brownian motion model, which does not preserve a continuous curvature. The crawling creatures and wobbling robots are modeled as active directional filaments, being composed of otherwise identical body segments with the sole proviso, that only the head segment is active with its curvature undergoing a steered active Ornstein-Uhlenbeck stochastic process. The rest of the segmented body is assumed to passively trail after the active head so that its dynamical state simply lags by various amounts from the state of the head segment.

We obtain the probability density of the active head by solving the appropriate Fokker-Planck equation and from this we calculate the probability density for the whole filament by integrating the probability density of the active head along the rest of the segmented filamentous body. The probability density of the whole filament is found to be partially heterogeneous at small time scales but becomes completely heterogeneous at large time scales with extended distribution of orientation and localized distribution of curvature.
The various correlators of the dynamical state of the segmented filament are calculated by three different methods, \emph{i.e.}, analytically from the Fokker-Planck equation, analytically from the Langevin equations, as well as numerically by integrating the Langevin equations.

The results obtained by three different methods coincide. Initial orientation and curvature are shown to be irrelevant at large time in the expression for the mean-square orientational fluctuations, the orientational correlation function and the mean-square head-to-tail separation. A characteristic orientation emerges, depending on the activity, diffusion of curvature and curvature time decay constant, which determines the mean-square orientational fluctuations and consequently the orientational correlation function and the mean-square head-to-tail separation. Our theoretical model sheds new light on the metameric locomotion of segmented animals but can also be useful in the design of metameric, segmented robots~\cite{Hoffman2013,calderon2016,agostinelli2018,zhan2019}.

\section{Acknowledgements}
R.P.~acknowledges the support of the Key project of the National Natural Science Foundation of China (NSFC) (Grant No.~12034019). F.Y.~acknowledges the support of the National Natural Science Foundation of China (NSFC) (Grant No.~11774394), the Strategic Priority Research Program of Chinese Academy of Sciences (Grant No.~XDB33030300), and the K.~C.~Wong Education Foundation.

\end{document}


\title{Model of metameric locomotion  in active directional filaments\\ \normalfont\textit{Supplementary Material}}


\maketitle

\section{Derivation of probability density from Fokker-Planck equation}

Here we solve the probability density \(P_{0}(\theta,\kappa,t | \theta_{0}, \kappa_{0}, 0)\) fulfilling the Fokker-Planck equation
 \begin{align}
\label{eq:fp}
 \frac{\partial }{\partial t} P_{0}
 = - \frac{\partial}{\partial\theta}\left(v_{0}\kappa P_{0}\right)
+ \frac{\partial}{\partial \kappa} \left(\beta \kappa P_0\right)
+ \frac{g}{2} \frac{\partial^{2}}{\partial \kappa^{2}} P_0
 \end{align}
with initial condition
\begin{align}
 P(\theta, \kappa,t| \theta_{0}, \kappa_{0},0)|_{t=0} = \delta(\theta-\theta_{0}) \delta(\kappa-\kappa_{0}).
\end{align}
However, we will not solve the above Fokker-Planck equation directly. We first observe the Langevin equations, from which the Fokker-Planck equation is derived
\begin{subequations} \label{eq:langevin}
 \begin{align}
 \frac{\mathrm{d}\theta(t)}{\mathrm{d}t} &= v_{0} \kappa(t), \\
 \frac{\mathrm{d}\kappa(t)}{\mathrm{d}t} &= -\beta \kappa(t) + \xi(t),
 \end{align}
\end{subequations}
where
\begin{align}
\langle \xi(t)\rangle = 0, \quad \langle\xi(t)\xi(t')\rangle = g \,\delta(t-t').
\end{align}

Rearrangement of Eqs.~\eqref{eq:langevin} into matrix form gives us
\begin{align}
\frac{\mathrm{d}}{\mathrm{d}t}
\begin{pmatrix}
\theta \\
\kappa
\end{pmatrix}
= -
\begin{pmatrix}
0 & -v_{0} \\
0 & \beta \\
\end{pmatrix}
\begin{pmatrix}
\theta \\
\kappa
\end{pmatrix}
+
\begin{pmatrix}
0 \\
\xi
\end{pmatrix},
\end{align}
which is just two-dimensional Ornstein-Uhlenbeck process.
Denote
\begin{align}
\mathbf{B} =
\begin{pmatrix}
0 & - v_{0} \\
0 & \beta
\end{pmatrix}, \quad
\mathbf{D} =
\begin{pmatrix}
0 & 0 \\
0 & \frac{g}{2}
\end{pmatrix}, \quad
\mathbf{z} =
\begin{pmatrix}
\theta \\
\kappa
\end{pmatrix}.
\end{align}
We have the Fokker-Planck equation for two-dimensional Ornstein-Uhlenbeck process
\begin{align} \label{eq:2d-ou}
\frac{\partial P}{\partial t} = B_{ij} \frac{\partial}{\partial z_{i}} (z_{j} P) + D_{ij} \frac{\partial^{2} P}{\partial z_{i}\partial z_{j}}.
\end{align}
The solution of Eq.~\eqref{eq:2d-ou} is well known~\cite{risken1989} and given by
\begin{align} \label{eq:prob-2d-ou}
& P(\theta, \kappa, t | \theta_{0}, \kappa_{0}, 0) = (2\pi)^{-1} (\text{det} \sigma)^{-1/2} \exp\left\{
  -\frac{1}{2} \left[\sigma^{-1}(t)\right]_{\theta\theta} \left[\theta-\theta(t)\right]^{2} \right.\nonumber\\
& \quad \left.  -\left[\sigma^{-1}(t)\right]_{\theta\kappa} \left[\theta-\theta(t)\right]\left[\kappa-\kappa(t)\right]
  -\frac{1}{2} \left[\sigma^{-1}(t)\right]_{\kappa\kappa} \left[\kappa-\kappa(t)\right]^{2}
\right\},
\end{align}
where
\begin{gather}
\begin{pmatrix}
\langle\theta\rangle \\
\langle\kappa\rangle
\end{pmatrix}
=
\begin{pmatrix}
\theta(t) \\
\kappa(t)
\end{pmatrix}
= G(t)
\begin{pmatrix}
\theta_{0} \\
\kappa_{0}
\end{pmatrix}, \quad
G(t) = \exp(-\mathbf{B}t),\nonumber\\
\sigma_{ij}(t) = \int_0^t G_{ik}(t') G_{jl}(t') \mathrm{d}t' 2D_{kl}.
\end{gather}

We now use spectral decomposition to write the above equations explicitly. We express \(\mathbf{B}\) in biorthogonal vectors (repeated indices are summed)
\begin{align}
& \mathbf{B} = \lambda_{\alpha} \mathbf{u}^{(\alpha)} \mathbf{v}^{(\alpha)}, \quad
\lambda_{1} = 0, \quad \lambda_{2} = \beta, \nonumber\\
& \mathbf{u}^{(1)} =
\begin{pmatrix}
1 \\ 0
\end{pmatrix}, \quad
\mathbf{u}^{(2)} =
\begin{pmatrix}
v_{0} \\ -\beta
\end{pmatrix}, \quad
\mathbf{v}^{(1)} =
\begin{pmatrix}
1 & v_{0}/\beta
\end{pmatrix}, \quad
\mathbf{v}^{(2)} =
\begin{pmatrix}
0 & -1/\beta
\end{pmatrix}.
\end{align}
Then
\begin{align}
G(t) = \exp(-\mathbf{B}t) = \mathrm{e}^{-\lambda_{\alpha}t} \mathbf{u}^{(\alpha)} \mathbf{v}^{(\alpha)} =
\begin{pmatrix}
1 & \frac{v_{0}}{\beta} \left(1-\mathrm{e}^{-\beta t}\right) \\
0 & \mathrm{e}^{-\beta t}
\end{pmatrix}.
\end{align}
Finally we obtain explicitly the mean orientation and curvature
\begin{align}
\theta(t) = \theta_0 + \frac{1-\mathrm{e}^{-\beta t}}{\beta} v_0\kappa_0, \quad \kappa(t) = \mathrm{e}^{-\beta t} \kappa_0,
\end{align}
and the variance matrix
\begin{align}
\sigma =
\begin{pmatrix}
\frac{gv_{0}^{2}}{2\beta^{3}} \left( 2\beta t - \mathrm{e}^{-2\beta t} + 4\mathrm{e}^{-\beta t} - 3\right) &
\frac{gv_{0}}{2\beta^{2}} \left( \mathrm{e}^{-\beta t} - 1\right)^{2} \\
\frac{gv_{0}}{2\beta^{2}} \left( \mathrm{e}^{-\beta t} - 1\right)^{2} &
\frac{g}{2\beta} \left(1 - \mathrm{e}^{-2\beta t}\right)
\end{pmatrix},
\end{align}
which completes the solution of the Fokker-Planck equation Eq.~\eqref{eq:fp}.

Now that we have the joint probability density dependent on both orientation and curvature, we can readily obtain the marginal
probability density dependent separately only on orientation and curvature.
The marginal distribution of multivariate normal distribution can be obtained by just dropping irrelevant parts, which can be easily proved.
Hence we obtain the following two marginal probability distributions
\begin{align}
P(\theta, t | \theta_{0}, 0) = (2\pi)^{-1/2} \sigma_{\theta\theta}^{-1/2} \exp\left\{ -\frac{1}{2} \sigma_{\theta\theta}^{-1} \left[\theta-\theta(t)\right]^{2}\right\}, \\
P(\kappa, t | \kappa_{0}, 0) = (2\pi)^{-1/2} \sigma_{\kappa\kappa}^{-1/2} \exp\left\{ -\frac{1}{2} \sigma_{\kappa\kappa}^{-1} \left[\kappa-\kappa(t)\right]^{2}\right\},
\end{align}
where
\begin{align}
\sigma_{\theta\theta} = \frac{gv_{0}^{2}}{2\beta^{3}} \left( 2\beta t - \mathrm{e}^{-2\beta t} + 4\mathrm{e}^{-\beta t} - 3\right), \quad
\sigma_{\kappa\kappa} =  \frac{g}{2\beta} \left(1 - \mathrm{e}^{-2\beta t}\right).
\end{align}

\section{Derivation of probability density with orientationally steering field}

The presence of orientationally steering field can profoundly alter the behavior of the probability density of the active head at large time.
Assume the motion of the active head is orientationally steered by a harmonic potential
\begin{align}
  V = \frac{1}{2} \epsilon (\theta-\theta_{\mathrm{p}})^{2},
\end{align}
where \(\epsilon\) is a positive constant and \(\theta_{\mathrm{p}}\) represents the preferred orientation.
Then the Langenvin equations become
\begin{align}
\frac{\mathrm{d}\theta(t)}{\mathrm{d}t} &= v_{0} \kappa(t),\\
\frac{\mathrm{d}\kappa(t)}{\mathrm{d}t} &= -\beta \kappa(t) + \xi - \epsilon(\theta-\theta_{\mathrm{p}}).
\end{align}
Denote \(\lambda_{1,2}= \frac{1}{2}(\beta\pm\sqrt{\beta^{2} - 4\epsilon v_{0}})\).
The solution of corresponding Fokker-Planck equation is like Eq.~\eqref{eq:prob-2d-ou}
\begin{align}
& P'_{0}(\theta, \kappa, t | \theta_{0}, \kappa_{0}, 0) = (2\pi)^{-1} (\text{det} \sigma')^{-1/2} \exp\left\{
  -\frac{1}{2} \left[\sigma'^{-1}(t)\right]_{\theta\theta} \left[\theta-\theta'(t)\right]^{2} \right.\nonumber\\
& \quad \left.  -\left[\sigma'^{-1}(t)\right]_{\theta\kappa} \left[\theta-\theta'(t)\right]\left[\kappa-\kappa'(t)\right]
  -\frac{1}{2} \left[\sigma'^{-1}(t)\right]_{\kappa\kappa} \left[\kappa-\kappa'(t)\right]^{2}
\right\},
\end{align}
where \(\theta'(t)\) and \(\kappa'(t)\) are, respectively, the mean orientation and curvature
\begin{align}
\theta'(t) &= \frac{1}{\lambda_{1}-\lambda_{2}} \left[(-\lambda_{2}\mathrm{e}^{-\lambda_{1}t} + \lambda_{1}\mathrm{e}^{-\lambda_{2}t})\theta_{0} + v_{0}(-\mathrm{e}^{-\lambda_{1}t} + \mathrm{e}^{-\lambda_{2}t}) \kappa_{0}\right], \\
\kappa'(t) &= \frac{1}{\lambda_{1}-\lambda_{2}} \left[\epsilon(\mathrm{e}^{-\lambda_{1}t} - \mathrm{e}^{-\lambda_{2} t})\theta_{0} + (\lambda_{1}\mathrm{e}^{-\lambda_{1}t} - \lambda_{2}\mathrm{e}^{-\lambda_{2}t}) \kappa_{0}\right],
\end{align}
and \(\sigma'\) is the variance matrix with elements being
\begin{subequations}
\begin{align}
  \sigma'_{\theta\theta} &= \frac{v_{0}^{2} (g+\epsilon^{2}\theta^{2}_{\mathrm{p}})}{2(\lambda_{1}-\lambda_{2})^{2}}
 \left\{ \frac{\lambda_{1}+\lambda_{2}}{\lambda_{1}\lambda_{2}}
         + \frac{4}{\lambda_{1}+\lambda_{2}}\left[ \mathrm{e}^{-(\lambda_{1}+\lambda_{2})t}-1\right]
         - \frac{1}{\lambda_1} \mathrm{e}^{-2\lambda_1 t} - \frac{1}{\lambda_2} \mathrm{e}^{-2\lambda_2 t}
 \right\}, \\
 \sigma'_{\theta\kappa} &= \sigma'_{\kappa\theta} = \frac{v_0(g+\epsilon^2\theta^{2}_{\mathrm{p}})}{2(\lambda_1 - \lambda_2)^2} \left( \mathrm{e}^{-\lambda_1 t} - \mathrm{e}^{-\lambda_2 t}\right)^2, \\
 \sigma'_{\kappa\kappa} &= \frac{g+\epsilon^2\theta^{2}_{\mathrm{p}}}{2(\lambda_1 - \lambda_2)^2}
 \left\{
\lambda_1 + \lambda_2 + \frac{4\lambda_1 \lambda_2}{\lambda_1 + \lambda_2} \left[ \mathrm{e}^{-(\lambda_1 + \lambda_2)t} - 1\right]
- \lambda_1 \mathrm{e}^{-2\lambda_1 t} - \lambda_2 \mathrm{e}^{-2\lambda_2 t}  \right\}.
\end{align}
\end{subequations}
Note the real parts of \(\lambda_{1,2}\) are always positive.
Hence when \(t \to \infty\),
\begin{subequations}
\begin{align}
  & \theta'(t) \to 0, \quad \kappa'(t) \to 0, \\
  & \sigma'_{\theta\theta} \to \frac{v_{0}(g+\epsilon^{2}\theta_{\mathrm{p}}^{2})}{2\beta\epsilon}, \quad \sigma'_{\theta\kappa}=\sigma'_{\kappa\theta} \to 0,
  \quad \sigma'_{\kappa\kappa} \to \frac{g+\epsilon^{2}\theta_{\mathrm{p}}^{2}}{2\beta}.
\end{align}
\end{subequations}
The presence of orientationally steering field renders the mean orientation vanish and the variance of orientation finite at large time compared to the case without orientationally steering field.
It is noteworthy that the transition of orientational variance from diverging to finite is continuous at \(\epsilon=0\).

\section{Filament probability density and correlators at small time}

In the main text, we focus on the large time regime with \(t > T\), where \(T\) is
the time for the filament to move a distance equal to its body length. We now discuss the marginal case with \(t\leq T\).
A metameric animal may drop accidentally into a new environment or a metameric robot can be set manually to start moving.
When \(t \leq T\), part of the distal segments remain at the initial state, having not become the lagged state of the active head yet.

Assume the initial state of the filament is specified by the position \(\mathbf{r}_{\tilde{l}}(0)\), orientation \(\theta_{\tilde{l}}(0)\) and curvature \(\kappa_{\tilde{l}}(0)\).
For \(0 < \tilde{l} \leq t/T\), \(\theta_{\tilde{l}}(t) = \theta_0(t- \tilde{l}T)\) and \(\kappa_{\tilde{l}}(t) = \kappa_0(t- \tilde{l}T)\);
for \(t/T < \tilde{l} \leq 1\), \(\theta_{\tilde{l}}(t) = \theta_{\tilde{l}}(0)\) and \(\kappa_{\tilde{l}}(t) = \kappa_{\tilde{l}}(0)\).
The probability density of finding a segment of a single filament with orientation \(\theta\) and curvature \(\kappa\) at time \(t \leq T\) is
\begin{align}
  P(\theta, \kappa, t)  =& \int_{0}^{1} \mathrm{d}\tilde{l} \langle \delta(\theta-\theta_{\tilde{l}}(t)) \delta(\kappa-\kappa_{\tilde{l}}(t))\rangle_{\xi} \nonumber\\
  =&\, \frac{1}{T}\int_{0}^{t}\mathrm{d} t' P_{0}(\theta, \kappa, t') + \int_{t/T}^{1} \mathrm{d}\tilde{l}\, \delta(\theta-\theta_{\tilde{l}}(t)) \delta(\kappa-\kappa_{\tilde{l}}(t)).
\end{align}
The mean-square separation between the head and tail is
\begin{align}
  & \langle(\mathbf{r}_{0}(t) - \mathbf{r}_{1}(t))^2\rangle_{\xi} = \langle(\mathbf{r}_{0}(t) - \mathbf{r}_{1}(0))^2\rangle_{\xi} \nonumber \\
  =& \langle(\mathbf{r}_{0}(t) - \mathbf{r}_{0}(0) + \mathbf{r}_{0}(0) -  \mathbf{r}_{1}(0))^2\rangle_{\xi}\nonumber \\
  =& \langle(\mathbf{r}_{0}(t) - \mathbf{r}_{0}(0) )^2\rangle_{\xi} + 2\langle\mathbf{r}_{0}(t) - \mathbf{r}_{0}(0) \rangle_{\xi}\cdot(\mathbf{r}_{0}(0) -  \mathbf{r}_{1}(0))+ (\mathbf{r}_{0}(0) -  \mathbf{r}_{1}(0))^2
\end{align}
We have calculated the term \(\langle\left(\mathbf{r}_{0}(t)-\mathbf{r}_{0}(0)\right)^{2}\rangle_{\xi}\) in the main text.
\(\langle\mathbf{r}_{0}(t)-\mathbf{r}_{0}(0)\rangle_{\xi}\) can be calculated by
 \begin{align}
\langle\mathbf{r}_{0}(t)-\mathbf{r}_{0}(0)\rangle_{\xi} = v_{0} \int_{0}^{t} \mathrm{d}t_{1} \langle\mathbf{t}(t_{1})\rangle_{\xi},
\end{align}
where \(\langle\mathbf{t}(t)\rangle_\xi\) again can be calculated by \(\langle \theta(t)\rangle_{\xi}\) and \(\langle\theta^2(t)\rangle_{\xi}\), which are equivalently
\(\langle\theta(t)-\theta_0\rangle_\xi\) and \(\langle(\theta(t) - \theta_0)^2\rangle_{\xi}\)
\begin{align}
  \langle\mathbf{t}(t)\rangle_\xi  = \exp\left[-\frac{1}{2}\left(\langle\theta^2(t)\rangle_\xi - \langle\theta(t)\rangle_\xi^2\right)\right]
  \left(\cos\langle\theta(t)\rangle_\xi, \sin\langle\theta(t)\rangle_\xi\right).
\end{align}
The orientational correlation function between the head and tail is
\begin{align}
  & \langle \mathbf{t}_0(t)\cdot\mathbf{t}_1(t)\rangle_\xi = \langle\cos(\theta(t)-\theta_1(0))\rangle_\xi \nonumber \\
  =& \exp\left\{-\frac{1}{2}\left[ \langle(\theta(t)-\theta_1(0))^2\rangle_\xi - \langle\theta(t)-\theta_1(0)\rangle_\xi^2 \right]\right\}
  \cos\langle\theta(t)-\theta_1(0)\rangle_\xi.
\end{align}
The mean and the mean-square orientational fluctuations between the head and tail
\(\langle(\theta(t)-\theta_1(0))\rangle_\xi\) and \(\langle(\theta(t)-\theta_1(0))\rangle_\xi^2\) are again trivially related to
\(\langle\theta(t)-\theta_0\rangle_\xi\) and \(\langle(\theta(t) - \theta_0)^2\rangle_{\xi}\).
Therefore, the calculations of the mean-square orientational fluctuations, the orientational correlation function and the mean-square separation between the head
and tail at time \(t\leq T\) are all reduced to the calculations of
\(\langle\theta(t)-\theta_0\rangle_\xi\) and \(\langle(\theta(t) - \theta_0)^2\rangle_{\xi}\).
From the main text, we know
\begin{align}
  \langle \theta(t) - \theta_0\rangle_\xi &= \frac{v_0\kappa_0}{\beta} (1-\mathrm{e}^{-\beta t}), \\
\langle(\theta(t) - \theta_{0})^{2}\rangle_{\xi}
&=\, \frac{gv_{0}^{2}}{\beta^{3}} \left(\beta t - 1 + \mathrm{e}^{-\beta t}\right)
+ \frac{v_{0}^{2}(\kappa_{0}^{2}\beta -g/2)}{\beta^{3}} \left(\mathrm{e}^{-\beta t} - 1\right)^2,
\end{align}
which completes the calculations of the correlators in the marginal case with \(t \leq T\).